\documentclass[aps,prb,twocolumn,showpacs,groupedaddress]{revtex4}
\usepackage[american]{babel}
\usepackage{color,graphicx,amsmath,amssymb,bm}

\begin{document}
\newcommand\ket[1]{\left|#1\right\rangle}
\newcommand\bra[1]{\left\langle#1\right|}
\newcommand\lavg{\left\langle}
\newcommand\ravg{\right\rangle}

\title{Decoherence in a Cooper pair shuttle}

\author{Alessandro Romito}
\author{Francesco Plastina}
\author{Rosario Fazio}

\affiliation{NEST-INFM \& Scuola Normale
         Superiore, I-56126 Pisa, Italy}

\date{\today}

\begin{abstract}
We examine decoherence effects in the Josephson current of a
Cooper pair shuttle. Dephasing due to gate voltage fluctuations
can either suppress or enhance the critical current
and also change its sign. The current noise spectrum displays a
peak at the Josephson coupling energy and shows a phase
dependence.

\end{abstract}

\maketitle

The Josephson effect~\cite{josephson62} consists in a
dissipation-less current between two superconducting electrodes
connected through a weak link~\cite{tinkham96,barone82}. The
origin of the effect stems from the macroscopic coherence of the
superconducting condensate. Since its discovery in 1962, the
research on devices based on the Josephson effect has been
achieving a number of important breakthroughs both in
pure~\cite{tinkham96} and applied physics~\cite{barone82}. One of
the most recent exciting developments is probably the
implementation of superconducting nano-circuits for quantum
information processing~\cite{makhlin01}, which requires the
ability to coherently manipulate these devices. By now, this has
been shown in several experiments in systems of small Josephson
junctions~\cite{qcomp}.

\begin{figure}[hb]
\begin{center}
\includegraphics[width=70mm]{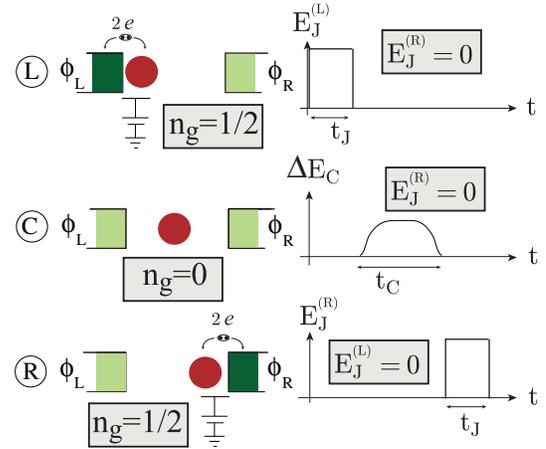}
\end{center}
\caption{Time dependence of the Josephson and charging energies in
the Cooper pair shuttle.The three intervals L, C and R, within
the period $T=2t_J+2t_C$, correspond to the situations: \textbf{L}) 
represents ${E_J}_L =E_J$, $n_g=1/2$, ${E_J}^{(R)} =0$ 
(interaction time); \textbf{C}) represents
${E_J}^{(L)} =0$, $n_g =0 $, ${E_J}^{(R)} =0$ (free evolution
time); \textbf{R}) ${E_J}^{(L)} =0$, $n_g=1/2$, ${E_J}_R =E_J$ -
(interaction time). On the left hand side, the corresponding
position of the shuttle with respect to the leads is shown for
each time interval.} \label{evolution}
\end{figure}

Very recently, Gorelik {\em et al.}~\cite{gorelik01,isacsson02}
proposed a very appealing setup, the Cooper pair shuttle, able to
create and maintain phase coherence between  two distant
superconductors. In its simplest realization, shown in
Fig.\ref{evolution}, the system is made up of a superconducting
grain, externally forced to move periodically between two
superconducting electrodes. Despite the fact that the grain is in
contact with only one lead at a time, the shuttle does not only
carry charge, as in the normal metal
case~\cite{gorelik98,erbe01,klein97}, but it also establishes
phase coherence between the superconductors.

Aim of this work is to analyze how the presence of the environment
affects the coherent transport in the Cooper pair shuttle. The
interplay~\cite{grifoni98} between the periodic driving and the
environmental dephasing leads to several interesting results.
By increasing the coupling to the
environment it may result in an {\em enhancement} of the supercurrent
as well as in a change of its sign ($\pi$-junction). In the last
part of this Letter we propose an effective implementation of the
shuttle mechanism where the switching of the Josephson couplings
is controlled by an external magnetic field.
The shuttle consists of a small superconducting island coupled to
two macroscopic leads and forced to change its position
periodically in time, with period $T$, from the Right ($R$) to the
Left ($L$) electrode and back (see Fig.\ref{evolution}). 
The grain
is small enough so that charging effects are important, while the
two leads are macroscopic and have definite phases $\phi_{L,R}$.
The moving island is described by the Hamiltonian
\begin{equation}
H_0 = E_C [\hat{n}- n_g(t)]^2 - \sum_{b =L,R} E_J^{(b)} (t)
\cos(\hat{\phi}-\phi_b)
\label{h0}
\end{equation}
where $E_C$ is the charging energy, $E_J^{(L,R)}$ is the Josephson
coupling to the left or right lead respectively, and $n_g$ is the
gate charge. The variable $\hat{n}$ is the number of excess Cooper
pairs in the grain and $\hat{\varphi}$ is its conjugate phase, $[
\hat{n}, \hat{\varphi} ]= -i$. The system operates in the Coulomb
blockade regime, ${E_J}^{(b)} \ll E_C$ ($b=R,L$) so that only the
two charge states $\{\ket{n=0}, \ket{n=1} \}$ are important. Their
relative energy
($\Delta E_C$) 
is controlled by the gate charge $n_g(t)$. The
superconducting gap is assumed to be the largest energy scale in
the problem, so that quasi-particle tunnelling can be neglected.
Both $E_C$ and $E_J^{(L,R)}$ are time dependent: when the grain is
close to one of the leads, the corresponding Josephson coupling is
non-zero (with value $E_J$) and the two charge states are
degenerate (positions $L$ and $R$ in Fig.\ref{evolution}). In the
intermediate region (position $C$), $E_J^{(L)}=E_J^{(R)}=0$. 
As in Ref. \cite{isacsson02} we employ a sudden approximation
(which requires a  switching time $\Delta t \ll 1/E_J$) and
suppose $E_J^{(L,R)}(t)$ to be step functions in each region (see
Fig.1). We further assume, as in Ref.\cite{gorelik01}, that the
system is at the charge degeneracy as long as the island is in
contact with one of the electrodes. 
In the intermediate region (C) it is not necessary to specify the 
exact variation of $n_g(t)$, only the time integral of the energy 
difference between the two charge states will enter in the results.

The shuttle is coupled via the charge operator $\hat{n}$ to an
environment described by the Caldeira--Leggett model~\cite{weiss99}
\begin{equation}
H_{int} = \hat{n} \sum_i \lambda_i (a_i+ a_i^{\dag})+H_{bath} \; .
\label{hi}
\end{equation}
In Eq.(\ref{hi}), $H_{bath}$ is the bath Hamiltonian, with boson
operators $a_i$, $a_i^{\dag}$ for its $i-th$ mode. The form of the
coupling in Eq.(\ref{hi}) can describe either gate voltage
fluctuations~\cite{makhlin01} or, in some limits, random switching
of background charges in the substrate~\cite{paladino02}.

In order to analyze the transport process, we evaluate the time
averaged supercurrent at steady state
\begin{equation}
I = \overline{\langle \hat{I}\rangle } \equiv
 \frac{1}{T} \int_0^T dt
\langle \hat{I}(t) \rangle \; ,
\label{current}
\end{equation}
and the power spectrum of the current fluctuations
\begin{equation}
S(\omega)=  \int_{-\infty}^{+\infty} d\! \tau \tilde{S}(\tau) e^{-i\omega
\tau}
\label{spectrum}
\end{equation}
where
\begin{equation}
\tilde{S}(\tau)=  \frac{1}{2}\overline{\langle \left[ \hat{I}
(t+\tau), \hat{I} (t) \right]_+\rangle}
-\overline{\langle \hat{I}(t+\tau)\rangle \langle
\hat{I}(t)\rangle} \; .
\label{spectrum2}
\end{equation}
In the Schr\"odinger picture, the current operator is ($\hbar=1$)
\begin{equation}
\hat{I}(t) =
 2e  E_J  \sin \left( \hat{\varphi} -\phi_L \right) \, \Theta_L (t)
\end{equation}
corresponding to the exchange of Cooper pairs with the left lead.
The function $\Theta_L (t)$ is defined as follows: $\Theta_L (t)=
1$ when the grain is in the $L$ region, and $\Theta_L(t) =0$
otherwise (the functions $\Theta_R(t)$ and $\Theta_C(t)$ 
are defined analogously). In order to
evaluate  Eqs.(\ref{current}, \ref{spectrum}), we need to compute
the reduced density matrix of the grain $\rho (t)$. After one
period the evolution of $\rho (t)$ can be computed through a map
$\mathcal{M}_t$ defined by $\rho (t+T) = \mathcal{M}_t \rho (t)$.
The stationary limit is obtained by studying the fixed point of
$\mathcal{M}_t$ ~\cite{shytov01}. Since only two charge states in
the grain are relevant, the reduced density matrix can be
parametrized, in the charge basis, as $\rho(t) = 1/2 \left[ {\rm 1} 
\hspace{-1.1mm} {\rm I}
\hspace{0.5mm} +\vec{\sigma} \cdot \vec{r}(t) \right]$,
where $\sigma_i$ ($i=x,y,z$) are the Pauli matrices and $r_i(t)=
\lavg \sigma_i \ravg$. The assumption of a stepwise varying
Hamiltonian
considerably simplifies the form of the map ${\mathcal{M}_t}$,
obtained as a composition of the time evolutions of $\rho$ in the
intervals L,C,R (see Fig. \ref{evolution}). Separately for each of
these intervals, the master equation for $\vec r (t)$ has the
form~\cite{cohen}
\begin{equation}
\dot{\vec{r}}(t) = \sum_{k \in \{L,R,C \}} \left[  G_k (t)
\vec{r}(t) + 2 \gamma_J(T_b) \vec{w}_k \right] \, \Theta_k (t)
\label{master}
\end{equation}
with $\vec{w}^{\dag}_{L,R} =
\tanh (E_J /T_b)
\begin{pmatrix}
\cos \phi_{L,R},&
\!\sin\phi_{L,R}, & \! 0 \end{pmatrix},
\vec{w}_{C}^{\dag}=\begin{pmatrix}  0, & \! 0,& \!0
\end{pmatrix}$,
\begin{equation}
G_{L,R} =
\begin{pmatrix}
-2 \gamma_J(T_b) & 0 & -E_J \sin\phi_{L,R} \cr 0 & -2
\gamma_J(T_b) & -E_J \cos\phi_{L,R} \cr E_J \sin\phi_{L,R} & E_J
\cos\phi_{L,R} & 0
\end{pmatrix} \; ,
\end{equation}
\begin{equation}
G_C = \begin{pmatrix}
 - \gamma_C(T_b) & - E_C & 0 \cr
 E_C &  - \gamma_C(T_b) & 0 \cr
 0 & 0 & 1
 \end{pmatrix}\; ,
\end{equation}
where the bath is taken in thermal equilibrium at temperature
$T_b$.

\noindent Here, $\gamma_J(T_b)$ and $\gamma_C(T_b)$ are the
temperature-dependent dephasing rates in the regions L, R and C,
respectively, obtained in the Born-Markov approximation. As an
example, for an ohmic bath with coupling to the environment 
$\alpha \ll 1$, one has~\cite{weiss99} 
$\gamma_J(T_b)= (\pi/2) \alpha E_J \coth (E_J/2T_b) $ and
$\gamma_C(T_b)= 2 \pi \alpha T_b$. This treatment is valid
provided that $\gamma_{J,C} \ll T_b, E_J$, and that the time
interval $t_{J(C)}$ is much longer than both $T_b^{-1}$ and
$E_{J(C)}^{-1}$.
In the coupling regions L and R, the only energy scale is set by
the Josephson energy, while, during the free evolution time, the
relevant scale is the energy difference between charge states.
Correspondingly, all the physical quantities depend on the phases
$2 \theta = E_Jt_J$ and $2 \chi = \int_C dt \Delta E_C (t)$
\cite{footnote1}. The other important variable is the phase
difference $\phi = \phi_L -\phi_R$. The effect of damping is
characterized by the two dimensionless quantities $\gamma_J t_J$
and $\gamma_C t_C$ \cite{footnote2}.

\noindent \underline{\emph{Average current}}- In the limiting case
considered by Gorelik {\em et al.}~\cite{gorelik01}, the Josephson
current does not depend on the dephasing rates. One expects,
however, that this cannot be always the case. If, for example, the
period T is much larger than the inverse dephasing rates, the
shuttle mechanism is expected to be inefficient and the critical
current should be strongly suppressed. In fact, we find a quite
rich scenario, depending on the relative value of the various time
scales and phase shifts.

\begin{figure}
\includegraphics[width=70mm]{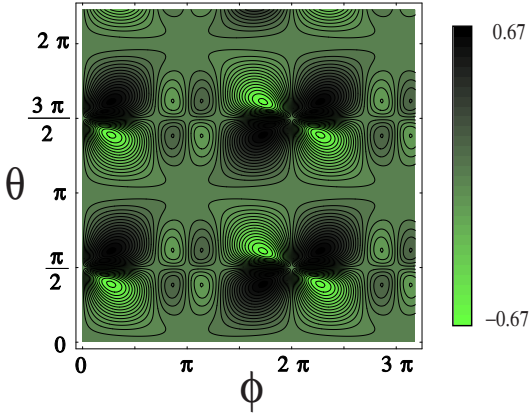}
\caption{Supercurrent (in units of $e/T$) as a function of the
superconductor phase difference $\phi$ and of the phase
accumulated during the contact to one of the electrodes $\theta$.
The other parameters are fixed as: $\chi=5 \pi/6$, $e^{-\gamma_J
t_J}=3/4$, $e^{-\gamma_C t_C}=4/5$. The plot is obtained for $T_b
\ll E_J$} \label{corrente1}
\end{figure}

\begin{figure}
\includegraphics[width=70mm]{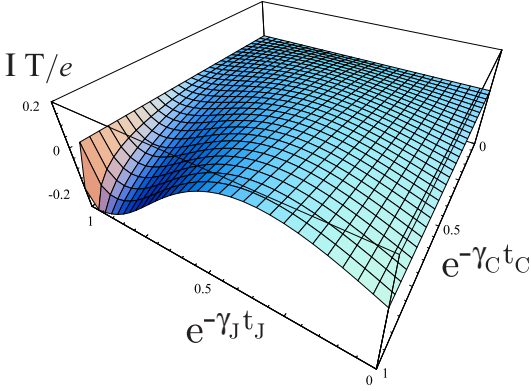}
\caption{Average current ($T_b \ll E_J$) as a function of the dephasing
rates, with $\phi=-3\pi /4$, $\theta=7 \pi /10$, $\chi=5 \pi/6$. As a
function of $\gamma_J t_J$, the supercurrent has a not-monotonous
behavior. Note the change of sign in the current obtained by
varying decoherence rates in each time interval separately.}
\label{corrente2}
\end{figure}

The expression of the current $I(\phi,\theta,\chi,\gamma _Jt_J,
\gamma _C t_C)$ can be obtained analytically from
Eqs.(\ref{current},\ref{master}). However, it is rather cumbersome
and not instructive, so we prefer to present it in some limiting
cases. A typical plot of $I$ as a function of $\theta$ and $\phi$
is shown in Fig.\ref{corrente1}. Depending on the value of
$\theta$ (a similar behaviour is observed as a function of
$\chi$), the critical current can be negative, i.e. the system can
behave as a $\pi$-junction. The phase shifts accumulated in the
time intervals L,C and R, leading to the current-phase relation
shown in Fig.\ref{corrente1}, are affected by the dephasing rates
in a complicated way. By changing $\gamma_J t_J$ and $\gamma_C
t_C$, certain interference paths are suppressed, resulting in a
shift of the interference pattern and ultimately in a change of
the sign of the current, as shown in Fig.\ref{corrente2}.

An analysis of the critical current as a function of the dephasing
rates reveals another interesting aspect: the Josephson current is
a non-monotonous function of $\gamma_J t_J$, i.e. by {\em
increasing} the damping, the Josephson current can {\em increase}.
The behavior as a function of the dephasing rates is
presented in  Fig.\ref{corrente2}. The presence of a maximum
Josephson current at a finite value of $\gamma_J t_J$ can be
understood by analyzing the asymptotic behaviors in the strong and
weak damping limits, where simple analytic expressions are available
(in the following we do not explicitely write the temperature dependence 
in $\gamma_J$, $\gamma_C$).\\
i)If the dephasing is strong, $I$ can be expanded in powers of
$e^{-\gamma_J t_J}$ and $e^{-\gamma_C t_C}$ and, to leading order
\begin{equation}
I_{strong} \sim  \frac{2 e}{T}
\tanh \left( \frac{E_J}{T_b} \right)
e^{-(\gamma_J t_J+\gamma_C t_C)}
\cos(2\chi)\sin(2\theta)  \, \sin\phi \, .
\label{limit0}
\end{equation}
Strong dephasing is reflected in the simple (i.e. $\propto \sin
\phi$) current-phase relationship and in the exponential
suppression of the current itself.\\
ii)In the opposite limit of weak damping ($\gamma_J t_J \ll
\gamma_C t_C \ll 1 $~\cite{footnote2}),
\begin{equation}
I_{weak} \sim
\frac{2e}{T}
\tanh \left(\frac{E_J}{T_b} \right)
\frac{\gamma_J t_J}{\gamma_C t_C}
\frac{(\cos\phi+\cos 2\chi)\tan\theta \sin\phi}
{1+\cos\phi \cos2\chi} \, .
\label{limit1}
\end{equation}
The current tends to zero if the coupling with the bath is
negligible during the interaction time. In this case, indeed, the
time evolution in the intervals $L,R$ is almost unitary, while, in
the region $C$, pure dephasing leads to a suppression of the
off-diagonal terms of the reduced density matrix $\rho (t)$. As a
result, in the stationary limit the system is described by a
complete mixture with equal weights.
The current then tends to zero in both limiting cases of large and
small $\gamma_J t_J$. Therefore one should expect an optimal coupling
to the environment where the Josephson current is maximum.
A regime where the crossover between the strong and weak damping
cases can be described in simple terms is the limit $\gamma_C
\rightarrow 0$, for a fixed value of $\theta$. For example, at
$\theta=\pi /4$ the current reads
\begin{footnotesize}
\begin{eqnarray}
&  & I = \frac{2e}{T}
\tanh \left(\frac{E_J}{T_b} \right) \nonumber \\
& & \times \frac{2e^{-\gamma_J t_J}[2e^{-2\gamma_J t_J}\cos\phi +
(1+e^{-4\gamma_J t_J})\cos2\chi]\sin\phi}
{(1+e^{-2\gamma_J t_J})(1+ e^{-2\gamma_J t_J}\cos\phi \cos2\chi
+e^{-4\gamma_J t_J})}
\label{crossover}
\end{eqnarray}
\end{footnotesize}
In the limit of vanishing $\gamma_J t_J$, Eq.(\ref{crossover})
corresponds to the situation discussed in~\cite{gorelik01}.
Indeed, both expressions are independent of the dephasing rates.
The difference in the details of the current-phase(s) relationship
are due to the different environment.

In all the three cases presented here,
Eqs.(\ref{limit0},\ref{limit1},\ref{crossover}), the change of
sign of the current as a function of the phase shifts $\theta$ or
$\chi$ is evident.

\begin{figure}
\includegraphics[width=70mm]{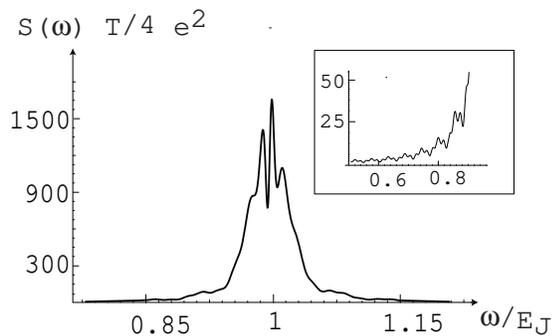}
\caption{Current noise spectrum as a function of $\omega$ for
$T=4t_J$  in strong dephasing limit
($\gamma_J t_J=1.2$ and $\gamma_J / E_J =0.008$). In the
inset, we plot the spectrum in a restricted range of frequencies
to better resolve the oscillations.} \label{noise}
\end{figure}

\noindent \underline{\emph{Current Noise}} - Cooper pair shuttling
is a result of a non equilibrium steady state process. Therefore,
to better characterize the transport, we analyze supercurrent
fluctuations as defined in Eq.(\ref{spectrum}). This should be
contrasted with the standard Josephson effect~\cite{tinkham96},
where the supercurrent is an equilibrium property of the system.
We first consider the zero frequency noise in the two regimes of 
strong and weak dephasing.\\
When the dephasing is strong, correlations on time scales larger than 
$T$ are suppressed and the noise spectrum reads
\begin{equation}
S(0)_{strong} \sim \frac{4 e^2}{T} \left\{ \frac{1}{2}-e^{-\gamma_J t_J}
\cos (2 \theta) + e^{-2\gamma_J t_J} f(\theta,\phi,\chi) \right\} \, ,
\label{strongnoise}
\end{equation}
where
$f(\theta, \phi, \chi) = \cos^2 (2\theta) -e^{-\gamma_C t_C} \cos \phi \cos 
\chi \sin^2 (2 \theta) $. 
The leading term in Eq.(\ref{strongnoise}) is due to the damped
oscillations in the contact regions (L,R). The phase dependent contribution
is exponentially suppressed since it comes from correlations
over times larger than the period.
For weak dephasing (same limits of Eq. (\ref{limit1})), we find
\begin{equation}
S(0)_{weak} \sim \frac{4e^2}{T} \frac{1}{\gamma_C t_C}
\frac{\tan^2\theta \sin^2 \phi}{2(1+ \cos \phi \cos 2 \chi)} \, ,
\label{Slimit1}
\end{equation}
which shows a much richer structure as a function of the phases
$\theta$ and $\chi$~\cite{footnote3}.
Finally, we briefly discuss the finite frequency spectrum in the
case of strong dephasing (see Fig.\ref{noise}). Superimposed to
the peak at the Josephson energy, there are oscillations of
frequency of the order of $T^{-1}$. The presence of these
oscillations is related to the periodicity of the island
motion. The modification of these fringes as a function of
the phases is a signature of the coherence in the Cooper pair
shuttle.

\begin{figure}
\includegraphics[width=70mm]{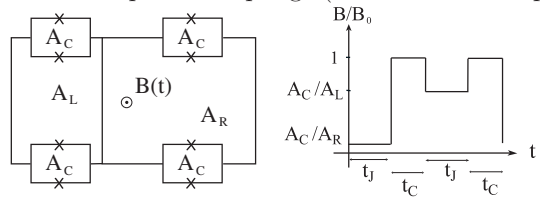}
\caption{
{\it left}: Sketch of the implementation of the shuttle process
by means of a time-dependent magnetic field. Crosses represent Josephson junctions.
{\it right}: Plot of the time variation of the applied field
(in unity of $B_0=\Phi_0/(2A_C)$,$\Phi_0$ is the flux quantum) in order to realize 
the Cooper pair shuttle. The different loop areas can be chosen in order to
obtain $E_J^{(L)}=E_J^{(R)}$.} 
\label{squid}
\end{figure}

We conclude by suggesting a  possible experimental test of our
results which does not require any mechanically moving part.
The time dependence of the Josephson couplings and $n_g$ is
regulated by a time dependent magnetic field and gate voltage,
respectively. The setup consists of a superconducting nanocircuit
in a uniform magnetic field as sketched in Fig.\ref{squid}. By
substituting the Josephson junction by SQUID loops, it is possible
to control the $E_J$ by tuning the applied magnetic field piercing
the loop. The presence of three type of loops with different area,
$A_L,A_R,A_C$ allows to achieve indipendently the three cases,
where one of the two $E_J$'s is zero (regions L,R) or both of them
are zero (region C), by means of a \emph{uniform} magnetic field.
If the applied field is such that a half-flux quantum pierces the
areas $A_L$,$A_R$ or $A_C$, the Josephson couplings will be those
of regions R,L and C, respectively and the Hamiltonian of the
system can be exactly mapped onto that of Eq.(\ref{h0}).
Moreover, by choosing the ratios $A_C/A_{R}=0.146$, and $A_C/A_L=0.292$ 
the two Josephson coupling are equal, $E_J^{(L)}=E_J^{(R)}$. 
This implementation has several advantages. It allows to control the 
coupling with the environment by simply varying the time dependence of the 
applied magnetic field. The time scale for the variation of the magnetic 
field should be controlled at the same level as it is done in the 
implementation of Josephson nanocircuits for quantum computation (see 
Ref.~\cite{makhlin01} for an extensive discussion). For 
a quantitative comparison with the results described here,
the magnetic field should vary on a time scale shorter than $\hbar / E_J$, 
tipically a few nanoseconds with the parameters of the first 
article in Ref.~\cite{qcomp}. This is possible with present day 
technology~\cite{buisson03}.
At a qualitive level the results presented in this paper ($\pi$-junction 
behavior, non-monotonous behavior in the damping) do not rely on the 
step-change approximation of the Josephson couplings (which leads to 
Eq.(\ref{master})). Those effects are observable even if 
the magnetic field changes on time-scales comparable or slower than $E_J$. 
The only strict requirement is that only one Josephson coupling at the 
time is switched on. 

We gratefully acknowledge many helpful discussions with G. Falci,
Yu. Galperin and Yu.V. Nazarov.
This work was supported by the EU (IST-SQUBIT, HPRN-CT-2002-00144)
and by Fondazione Silvio Tronchetti Provera.


\begin{thebibliography}{99}
\bibitem{footnote2}
        For some parameter values (e.g. $(\gamma_J,\gamma_C)=(0,0)$ or
        $(\gamma_J,\theta)=(0,\pi/2)$) our model of
        the environment is not effective; i.e. the system keeps memory of
        its initial conditions and the expressions given in the text
        cannot be used. Experimentally, other sources of dissipation are
        present which will matter for these parameter values.
\end{thebibliography}
\end{document}